# Fast Histograms using Adaptive CUDA Streams

Sisir Koppaka[*,1,2], Dheevatsa Mudigere[2], Srihari Narasimhan[2], Babu Narayanan[2]

[1]Depts. Of Mechanical & Industrial Engineering
Indian Institute of Technology, Kharagpur
Kharagpur, India

[2]Computing & Decision Sciences Lab
GE Global Research, JFWTC
Bangalore, India

*Abstract*— **Histograms are widely used in medical imaging, network intrusion detection, packet analysis and other stream-based high throughput applications. However, while porting such software stacks to the GPU, the computation of the histogram is a typical bottleneck primarily due to the large impact on kernel speed by atomic operations. In this work, we propose a stream-based model implemented in CUDA, using a new adaptive kernel that can be optimized based on latency hidden CPU compute. We also explore the tradeoffs of using the new kernel vis-à-vis the stock NVIDIA SDK kernel, and discuss an intelligent kernel switching method for the stream based on a degeneracy criterion that is adaptively computed from the input stream.**

## I. INTRODUCTION

In this paper, we address the problem of costly atomics in histogram computation as a part of various software stacks[1-3]. There have been several previous works in literature examining histogram computation on GPUs. NVIDIA[4] proposed a method for 256-bin histograms wherein a distinct sub-histogram is maintained per warp which gives it 9.07GBPS(*Giga Bytes Per Second*) throughput on the C1060. Shams et al[5] proposed two methods, the best of which performs at 11.3 GBPS. Their first method was based on Mark Harris's software-simulated atomics. Their second method was to do an update of the global memory only when the shared memory bin overflows. Their shared memory store was also bit-optimized. AMD[6] described a method utilizing OpenGL functions that can perform a 1024*1024 size data into a 256-bin histogram compute in 5.87ms.

However, in practice, most input data is either too large to fit into memory or online computation of histogram is required. Therefore, for large data streams, approximate frequency estimation techniques have been studied on the GPU[7]. Although CUDA streams have been available, they have not been studied in the context of exact histogram computation.

In addition, a realistic scenario of an input distribution peaking at a certain value will result in serializing a large number of atomics if computed on the GPU. Therefore, we study a stream-based model which utilizes CPU compute to optimize a proposed kernel, Adaptive Histogram Kernel (AHist), and to switch to a different kernel if necessary. We employ the NVIDIA CUDA programming model for our implementation.

## II. CUDA PROGRAMMING MODEL

NVIDIA's Compute Unified Device Architecture(CUDA), provides a general purpose-programming model for GPUs. This has increased the usage of GPU's for general purpose compute in addition to traditional visualization. The model for general purpose computation on the GPU is the GPU kernel. The GPU kernels are invoked from the CPU, offloading the compute intensive work onto the GPU. In the CUDA programming model, a kernel operates as a grid of thread-blocks.

Shared memory is a limited amount of very fast memory available on each SM assigned to a thread block. Global memory is far slower than shared memory and data read/write operations to global memory are costly and have to be coalesced for effective usage. However, a consistent issue with most GPUs and the CUDA programming model are the slow atomic operations on shared memory. For many practical tasks, like histogram computation, which are a frequent and recurring component of several algorithms the atomic operations end up serializing an enormous chunk of the compute resulting in a huge drop in performance. We examine here whether it is possible to use a different binning pattern in shared memory and additional CPU compute that could help in optimizing the GPU kernel performance by reducing the number of serialized atomics. A more detailed description of atomic operations supported by NVIDIA is provided in the CUDA programming guide [8].

*Corresponding author – sisir.koppaka@acm.org, Undergraduate student at IIT Kharagpur, work carried out at GE Global Research during May-July, 2010.

## III. COMPUTING HISTOGRAMS ON GPU

Our solution consists of an adaptive kernel that can adapt to different data streams – for example, we have presented below the study of this kernel for the case when it has adapted to normally distributed data similar to X-Ray image streams. The second part of the solution is to be able to run this kernel as a CUDA Stream with the CPU supplying the binning pattern appropriately. The third part is to intelligently decide whether the extra compute is necessary and switch the naïve NVHist and adaptive AHist kernels based on a criterion we define, called the "degeneracy" of the input distribution.

### *A. Adaptive Histogram Kernel (AHist)*

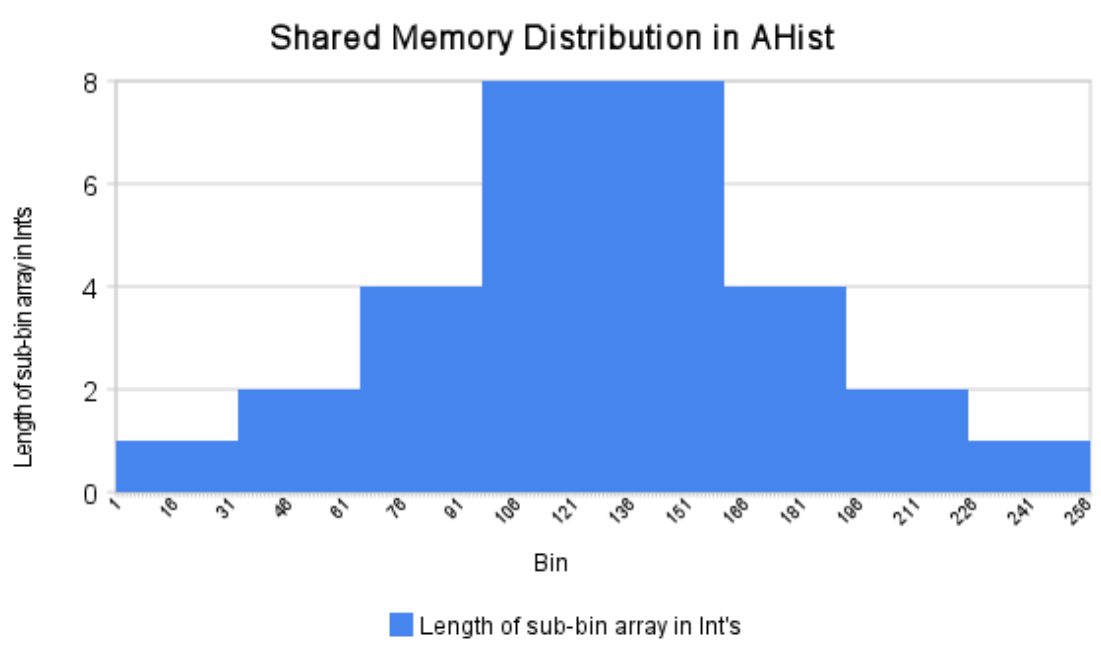

**Figure 1**

The AHist kernel reads 256 INTs which contain 1024 pixels (bytes) in a packed format, reads in another 256 INTs which form the indices for the bins in the sub-histogram in the shared memory of length 960(the binning pattern, shown in Figure 1), and writes out 256 INTs which form the partial histogram for the thread block. Each pixel is assumed to have a value between 0-255. The idea is to divide the atomics of a single bin in the histogram into multiple sub-bins to reduce their frequency of occurring (a divide-and-conquer approach). The genealogical performance of AHist on random input data is summarized in Table 1.

| Kernel | Description | Throughput (in GBPS) |
|---|---|---|
| 1 | Read, Write | 77.03 |
| 2 | 1+Initialize LocalHist=0 | 76.54 |
| 3 | 2 + Read binning pattern from global memory | 39.1 |
| 4 | 3 + Compute sub-histogram | 7.82 |
| 5 | 4+ Sum up per bin and write out | 6.89 |

**Table 1**

Several alternate ways were attempted at reducing the impact of step 3, i.e., reading the indices from global memory. These ways included hard-coding 8 different formulae through *if* conditions, 1 thread doing all the compute, half a warp doing all the compute, using constant memory, and passing it in a compressed form as a parameter to the kernel. The best performance was observed from reading it directly from global memory.

Although constant memory is faster than global memory, it's limited size and the improved throughput of coalesced global memory access patterns renders constant memory as a second tier choice in our case.

The second major drop occurs when we do atomicAdd. The extent of drop here is 5x. A similar drop of 10x is observed in the NVIDIA SDK kernel by virtue of introducing the atomicAdd. So the AHist kernel does help in reducing atomics, however, the overhead of obtaining the indices affects the impact of this reduction.

We compare the throughputs of AHist and the NVIDIA SDK Histogram (NVHist) in Table 2.

| Throughput on Input Data (in GBPS) | Random | Sequential | All equal to 127 | All equal to 1 | X-Ray Data |
|---|---|---|---|---|---|
| NVIDIA | 9.07 | 20.23 | 0.45 | 0.45 | 6.46 |
| AHist | 6.89 | 7.43 | 4.53 | 0.53 | 7.16 |

**Table 2**

The NVIDIA kernel performs faster than AHist for Random data and Sequential data. The overhead of additional calculations in AHist renders it slower NVIDIA for the Random case. This difference is particularly clear in the Sequential case, where AHist continues to suffer from additional calculations that are unnecessary for the given data, and NVIDIA benefits from a sequential pattern which speeds up it's atomics by reducing serialization and waiting times for a lock.

In the case when all values equal 127(for which we have maximum number of sub-bins and partitioning in shared memory), AHist performs very well obtaining a throughput of 4.53 GBPS versus 0.45 GBPS for NVIDIA. The 8 sub-bins reduce the number of atomic operations that slow down the kernel in this case for AHist. When all are equal to one, AHist is only very slightly ahead of NVIDIA. This slight difference can be attributed to the latency hiding obtained by AHist due to it's additional compute, since both AHist and NVIDIA have only 1 sub-bin for the first bin.

When the X-Ray data is supplied, AHist obtains a higher throughput than NVIDIA clearly showing the benefits of it's design.

The atomic add operations on sub-bins were minimized by allotting the sub-bins to each thread in a warp cyclically to minimize serialized writes within a warp. The same input slice data was chosen for both NVIDIA and AHist. It was fixed at 8192 X 8192 pixels, the default configuration of the NVIDIA kernel.

*B. Stream Histograms*

Dynamic histograms on streaming data are an important component of algorithms in several fields. For example, many computer vision and image processing algorithms rely on computing histogram-based objective functions with a sliding window. They are also applicable in approximating fast packet streams in routers for the purposes of intrusion detection and detecting Distributed-Denial-Of-Service attacks. The computational costs for these are high, and we develop a hybrid CPU-GPU approach to solve this problem.

Earlier, we noted that when all values were equal, AHist performed at 4.53 GB/s in comparison to the NVIDIA kernel at 0.45 GB/s for the region it was optimized for. This shows that AHist can be useful in distributions that possess temporal peaks. For example, as in a D-DOS attack on a router/network, or in detecting a marked change that requires reprocessing a slice in an Image Processing/Computer Vision context. However, the nature of the stream might change with time from say, normally distributed data, to another form. Therefore, we might have to intermittently recompute the indices for the sub-histogram, on the CPU, based on past stream's histograms. We found that computing the same on the GPU led to the kernel slowing down, and hence the present hybrid CPU-GPU approach was chosen.

For doing this, we construct two streaming histograms. The first is for an accumulated histogram, that gives us the general nature of the data stream. The second is a moving window histogram, which gives us the instantaneous nature of the data stream. By comparing them, we can detect if an attack is being mounted, or if there is a requirement for reprocessing a particular objective function in an image processing/computer vision algorithm.

The Accumulator, as we call it, implements a pipelined method for computing an accumulated histogram. The MW, implements a moving window histogram. We also implemented sequential versions of both to compare the pipeline performance.

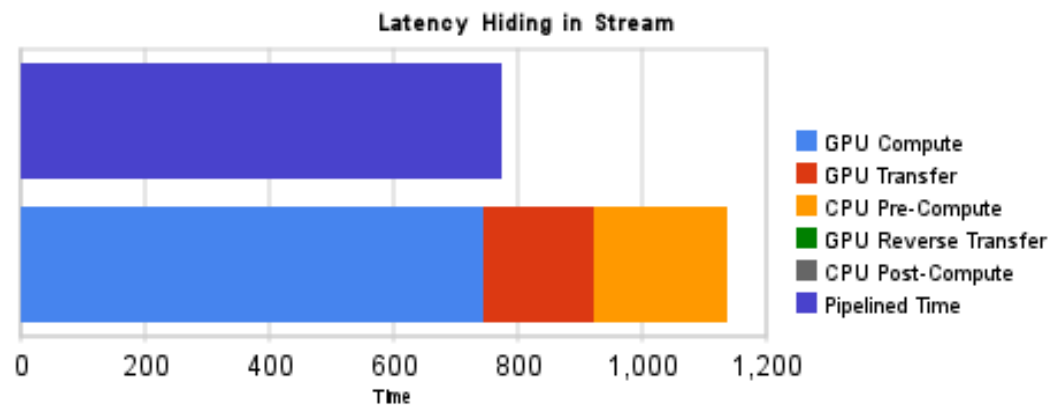

**Figure 2**

CUDA Streams are used for pipelining instructions to the GPU and CPU. CPU instructions like *memcpy* and recomputing the AHist indices are blocking, hence asynchronous instructions are relayed to the GPU and the CPU then carries out the blocking instructions for latency hiding, which is depicted in Figure 2 as a comparison between sequential streams and the latency hidden version.

The size of each chunk that is given to the GPU is decided by optimizing the balance between the compute time for the kernel and the PCI bandwidth manually. Sometimes, it is possible that an image slice, or a packet be too small and infrequent to demand an individual kernel invocation – in such cases multiple slices/packets can be batched within a single asynchronous call to return multiple histograms.

In order to prevent write conflicts due to the asynchronous nature of the stream, we use double buffering for all memory read/writes on the CPU & GPU for the stream, and employ at least one *cudaThreadSynchronize()* per iteration. This ensures latency hiding within each iteration, and prevents the stream from executing an iteration's kernel calls without fulfilling a previous dependent iteration's calls.

| 1 | 2 | 3 | 4 | 5 | 6 | 7 | 8 |
|---|---|---|---|---|---|---|---|
| R | 20.28 | 17.68 | 62.01 | 0.02 | 0 | 100 | 62.15 |
| S | 19.42 | 19.42 | 61.14 | 0.02 | 0 | 100 | 61.29 |
| N | 19.51 | 17.59 | 62.88 | 0.02 | 0 | 100 | 63.01 |

**Table 3**

1. Input data type, 2. CPU Pre-Compute, 3. GPU Transfer, 4. GPU Compute, 5. GPU Reverse Transfer, 6. CPU Post-Compute, 7. Total Sequential Time, 8. Pipelined Time, R-Random, S-Sequential, N- Normally Distributed Input data *Accumulator Histogram*

All values are percentages of Total Sequential Time

On the Accumulator, we note from Table 3 the benefit from pipelining, with the benefit being a 40% reduction in speed in comparison to a sequential version. Ideally, this would adapt to the state of the data stream, and normally distributed data is taken as an example here. The benefit from pipelining is also roughly proportional to number of streams to some extent, ranging from 97% for 1 stream, to 62% for 256 streams in general (shown in Figure 3).

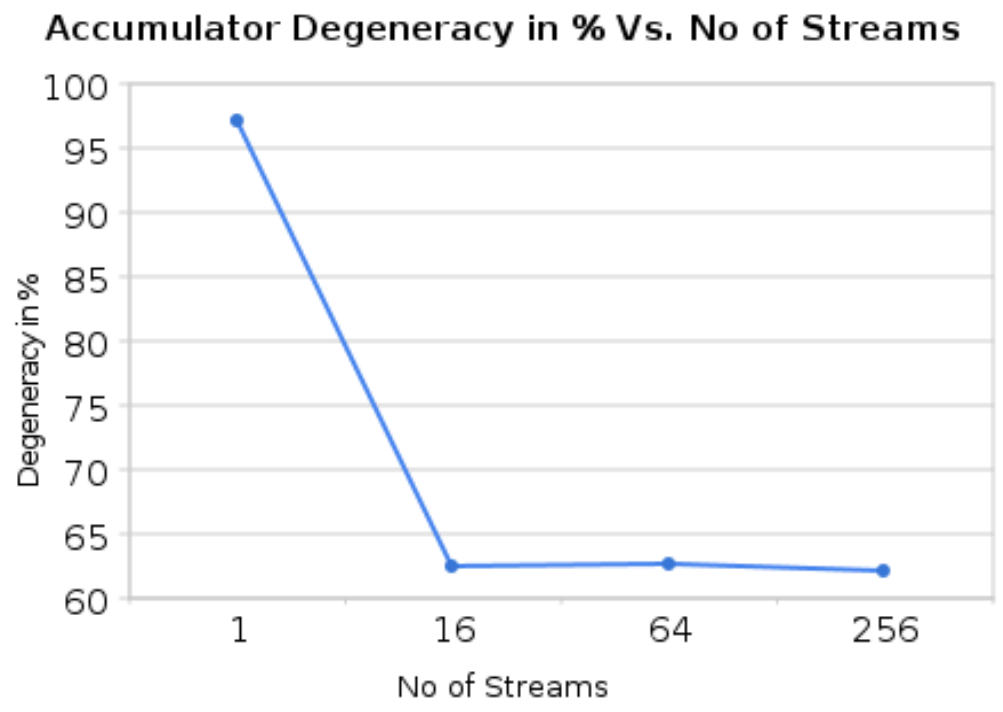

**Figure 3**

| 1 | 2 | 3 | 4 | 5 | 6 | 7 | 8 |
|---|---|---|---|---|---|---|---|
| 32 | 20.31 | 17.67 | 61.97 | 0.02 | 0.02 | 100 | 62.13 |
| 128 | 20.26 | 17.67 | 61.94 | 0.02 | 0.11 | 100 | 62.09 |
| 256 | 19.88 | 19.85 | 60.03 | 0.02 | 0.22 | 100 | 60.18 |

**Table 4**

1. Window Size, 2. CPU Pre-Compute, 3. GPU Transfer, 4. GPU Compute, 5. GPU Reverse Transfer, 6. CPU Post-Compute, 7. Total Sequential Time, 8. Pipelined Time, *Moving Window Histogram*

All values are percentages of Total Sequential Time

The MW stream algorithm was tested on varying window sizes(shown in Table 4) with the same random input data - implying that GPU Compute, GPU Transfer and GPU Reverse Transfer remain fairly constant, since the difference between the moving window and the accumulator is in the CPU post-compute part.

The benefit from pipelining here was also noticed to be proportional to the number of streams, ranging from 97% for 1 stream, to 61% for 256 streams (shown in Figure 4). This benefit from pipelining reduces slightly, or remains constant for an increase in number of windows for a given number of streams.

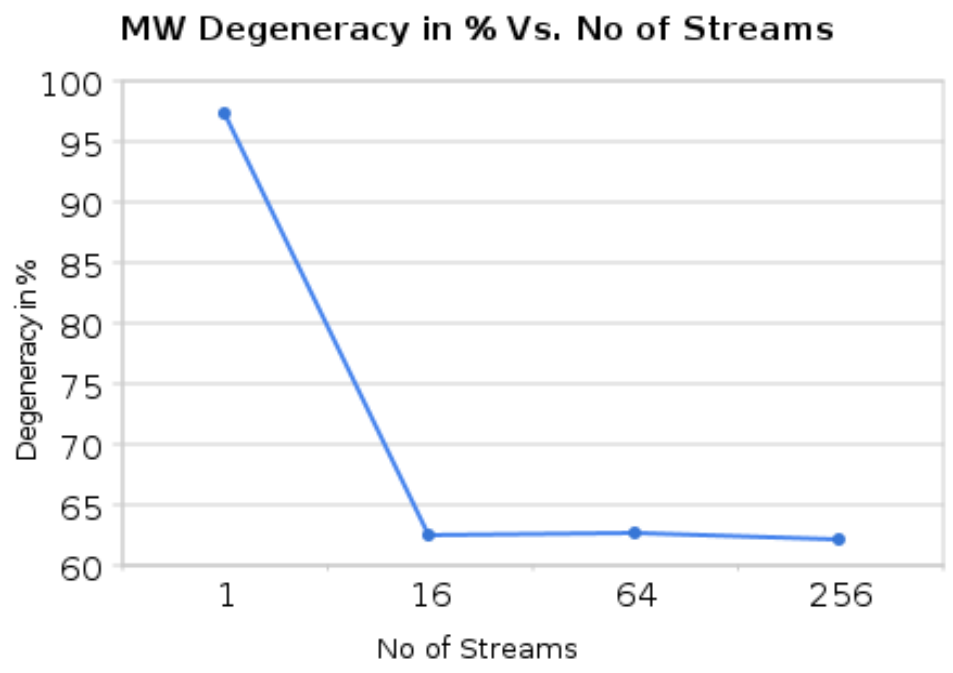

**Figure 4**

*C. Intelligent Kernel Switching*

To study the effect of atomics on the NVIDIA SDK kernel (NVHist) and AHist, and to simulate the effects of a time-variant distribution, we consider the combination of a degenerate distribution, and a random homogeneous distribution. A degenerate distribution is one where all the values in the distribution are equal to a single value (or lie in the same bin). We consider a distribution that consists of a degenerate distribution and a random homogeneous distribution in different ratios. In our case, the degeneracy of a given input distribution described above is determined by the percentage that consists of the degenerate distribution. This new distribution gives us the effect of the presence of the large number of atomics with a measure of control – the degeneracy.

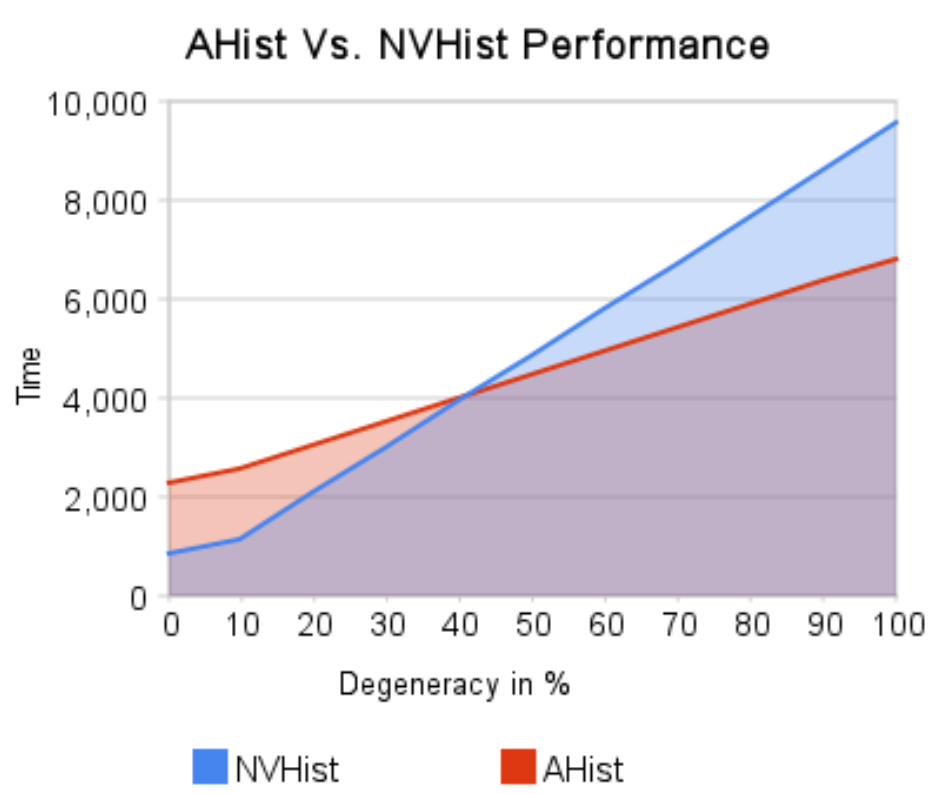

**Figure 5**

We note from Figure 5 that the extra compute in AHist is worth carrying out only given the cost of a sufficient number of atomic operations (denoted by degeneracy percentage). The streaming histogram implementation chooses between NVHist or AHist depending on the last computed moving window histogram, and evaluating if any of the bin values exceed the critical degeneracy required (between 40-50%).

## IV. CONCLUSION

We have described the three components of our solution, *viz.*, a new AHist kernel that can build a local histogram based on a supplied binning pattern, a streaming model which uses AHist after aggregating sufficient data from the stream and then invoking the kernel, and finally, intelligent kernel switching between NVHist and AHist in the stream based on the criterion of input distribution degeneracy determined online through latency hidden CPU compute on the MW histogram.

This solution can be used in situations where a CPU is usually available along with the GPU. We use the CPU to optimize a specific kernel (AHist), and also to switch to a different kernel (NVHist) once certain predetermined input distribution bounds are crossed. This feedback loop is implemented as a CUDA Stream, and latency hiding makes this model competitive to traditional one-kernel-does-it-all models. In practice, either most data is too large to fit in a GPU's memory, or online computation on the data is necessary. For both these cases, our model can be used effectively.

The possibility of switching criterion based on constraints other than degeneracy of the input distribution remains, and further studies are necessary to evaluate and compare various switching criterion. For example, combining degeneracy criterion with criterion that evaluate the amount of serialization possible by latency hidden rearrangement of data to the maximum extent possible can be explored.

## ACKNOWLEDGMENT

S.K. would like to thank Dr. Srikanth Rajagopalan, GE Global Research and JFWTC for giving him the opportunity to carry out this work at their facility.